\documentclass[prd,twocolumn,aps,nofootinbib,nobibnotes,superscriptaddress,preprintnumbers]{revtex4}

\usepackage{color}%to be able to use colors
\usepackage{amsmath}%Recomended for mathematical documents (use of: eqref,...)
\usepackage{graphicx}

\usepackage{color}%to be able to use colors
\usepackage{amsmath}%Recomended for mathematical documents (use of: eqref,...)
\usepackage{graphicx}

\newcommand{\dd}{{\rm d}}

\newcommand{\al}{\ensuremath\alpha}
\newcommand{\be}{\ensuremath\beta}

\newcommand{\Ga}{\ensuremath\Gamma}
\newcommand{\del}{\ensuremath\delta}
\newcommand{\la}{\ensuremath\lambda}

\newcommand{\cR}{\ensuremath\mathcal{R}}
\newcommand{\cT}{\ensuremath\mathcal{T}}
\newcommand{\cS}{\ensuremath\mathcal{S}}
\newcommand{\cU}{\ensuremath\mathcal{U}}

\newcommand{\beq}{\begin{equation}}
\newcommand{\eeq}{\end{equation}}

\newcommand{\na}{\nabla}

\newcommand{\lrsq}[1]{\left[#1\right]}

\newcommand{\sq}{\bar{q}}

\newcommand{\mq}{\hat{q}}
\newcommand{\mg}{\hat{g}}
\newcommand{\mT}{\hat{T}}

\renewcommand{\bf}[1]{{\textbf{#1}}}

\newcommand{\bL}{\ensuremath\tilde{L}}

\newcommand{\mpl}{M_{\rm Pl}}

\begin{document}

\title{Ghosts in metric-affine higher order curvature gravity}

\author{Jose Beltr\'an Jim\'enez}
\email[]{jose.beltran@usal.es}
\affiliation{Departamento de F\'isica Fundamental and IUFFyM, Universidad de Salamanca, E-37008 Salamanca, Spain.}

\author{Adria Delhom}
\email[]{adria.delhom@uv.es}
\affiliation{Departament de F\'{i}sica Te\`{o}rica and IFIC, Centro Mixto Universitat de
Val\`{e}ncia - CSIC.\\
Universitat de Val\`{e}ncia, Burjassot-46100, Val\`{e}ncia, Spain}

\begin{abstract}
We disprove the widespread belief that higher order curvature theories of gravity in the metric-affine formalism are generally ghost-free. This is clarified by considering a sub-class of theories constructed only with the Ricci tensor and showing that the non-projectively invariant sector propagates ghost-like degrees of freedom. We also explain how these pathologies can be avoided either by imposing a projective symmetry or additional constraints in the gravity sector. Our results put forward that higher order curvature gravity theories generally remain pathological in the metric-affine (and hybrid) formalisms and highlight the key importance of the projective symmetry and/or additional constraints for their physical viability and, by extension, of general metric-affine theories.
\end{abstract}

\maketitle

\section{Introduction}
Higher order curvature theories of gravity in the metric formalism exhibit pathologies caused by the higher order nature of their field equations that introduce Ostrogradski ghosts \cite{Woodard:2006nt,Woodard:2015zca}. This general statement is true for higher order curvature actions that contain arbitrary non-linear terms in the Riemann tensor, but there are special  theories that bypass the presence of these instabilities. On one hand, Lovelock theories are built out of non-linear combinations of the curvature with a special structure that guarantees having second order field equations and there are no additional propagating degrees of freedom of a potentially dangerous nature. Besides these, there are other ghost-free families of theories which do propagate additional modes, a paradigmatic example being $f(R)$ theories. They do give rise to fourth order field equations, but the Ostrogradski instability is avoided because the corresponding Hessian is degenerate. It is much more illuminating to analyse these theories in the Einstein frame where the healthy additional scalar field is apparent. Something similar happens for the functional extensions of the Gauss-Bonnet $f(\mathcal{G})$: No Ostrogradski instabilities are present there, although now a bit less trivially since the additional scalar field has Horndeski-type of interactions (see e.g. \cite{Kobayashi:2011nu}). These Horndeski interactions precisely conform the most general theory for a non-minimally coupled scalar field that give rise to second order field equations, thus avoiding Ostrogradski instabilities \cite{Horndeski} (see also \cite{beyondH}). Besides these exceptional cases, the presence of non-linear curvature interactions in the action will introduce additional ghostly modes.

It is broadly believed that the metric-affine formalism (also referred to as first-order or Palatini) avoids the pathologies associated to higher order curvature theories. The generally invoked reason is that the independence between the metric and the affine connection leads to second order field equations, so that one could naively expect to avoid Ostrogradski instabilities. The aim of this Letter is to clarify the incorrectness of this argument and to demonstrate the persistence of ghosts even in the metric-affine formalism and that care must be taken for higher order curvature theories regardless the employed formalism. Let us emphasise that, very much like having higher order field equations do not necessarily mean the presence of Ostrogradski instabilities as in the examples mentioned above, having second order field equations does not guarantee their absence. This is clear by noticing that any higher order theory can be recast in a second order form by introducing appropriate auxiliary fields. 

As a simplified proxy for higher order curvature gravity, we will consider actions depending only on the Ricci tensor, thus called Ricci-Based Gravity (RBG) theories. The motivation for this restriction is twofold: Firstly, the presence of ghosts can be readily shown, thus proving that higher order curvature theories are generically prone to instabilities. Secondly, these theories have received considerable attention, with some prevailing examples like the Eddington-inspired-Born-Infeld theory \cite{Banados:2010ix}, with its numerous extensions \cite{BIextensions} (see also \cite{BeltranJimenez:2017doy}) and the Ricci-square theories \cite{Riccisquare}. Although sometimes not explicitly stated, most of the literature on RBG further assumes projective symmetry in the gravity sector by imposing that only the symmetric part of the Ricci tensor contributes to the action. It is well-understood that these symmetric RBG theories do not propagate additional degrees of freedom (dof's) associated to the connection, and this fact can be traced back to the action having a projective symmetry. As a matter of fact, these theories are arguably nothing but General Relativity (GR) in disguise, since they admit an Einstein frame. This frame is achieved after integrating out the non-dynamical connection, whose effect is then to generate new interactions in the matter sector (see \cite{BeltranJimenez:2017doy,Afonso:2017bxr}). These matter interactions have in turn been used to place stringent constraints to symmetric RBG theories \cite{Latorre:2017uve}.

In this Letter we clarify what happens when the projective symmetry is explicitly broken. We show how these theories relate to the so-called Non-symmetric Gravity Theories (NGT) introduced by Moffat\footnote{Einstein had already considered non-symmetric metrics in an attempt to unify gravity and electromagnetism \cite{Einstein:1945eu}.} \cite{Moffat:1978tr} where the metric carries an antisymmetric part. These NGTs have been shown to exhibit certain pathologies \cite{Damour:1991ru,Damour:1992bt,Clayton:1996dz} that are then inherited by the general RBG. Moreover, we analyse in a more illuminating and manifest manner the presence of ghosts and Ostrogradski instabilities in RBG, what possess a serious drawback and signals the importance of the projective symmetry as a guide in the search for physically acceptable theories within a metric-affine approach. After properly identifying the ghosts in RBG, we show that constraining the connection to be torsion-free restores stability, with one extra massive vector field. Finally, we briefly discuss how the pathologies also transcend to the hybrid framework, thus showing the generic pathological nature of higher order curvature theories in any formalism, unless additional symmetries and/or restrictions are incorporated.

\section{Ricci-based metric-affine theories}
The RBG theories are described by
\beq\label{GeneralAction}
\cS[g_{\mu\nu},\Ga]=\frac12\int \dd^4x \sqrt{-g}\,F\big(g^{\mu\nu},\cR_{\mu\nu}(\Ga)\big)+\cS_{\rm m}[g_{\mu\nu},\Psi],
\eeq
where $F$ is any analytic scalar function depending on the inverse metric $g^{\mu\nu}$ and the Ricci tensor $\cR_{\mu\nu}$ of the connection $\Gamma^\alpha_{\mu\beta}$. The matter sector is assumed to be a collection of minimally coupled fields represented by $\Psi$. Unlike previous studies where only the symmetric part of the Ricci tensor was considered \cite{Banados:2010ix,BIextensions,Riccisquare}, we allow for its antisymmetric component as well, which explicitly breaks the projective symmetry\footnote{Under a projective transformation $\delta_\zeta\Gamma^\alpha{}_{\mu\beta}=\zeta_\mu \delta^\alpha_\beta$, the Riemann tensor changes as $\delta_\zeta \cR^\alpha{}_{\beta\mu\nu}=2\delta^\alpha{}_{\beta}\partial_{[\mu}\zeta_{\nu]}$ so that $\cR_{(\mu\nu)}$ remains invariant but $\cR_{\mu\nu}$ does not. The projective symmetry of RBG actions is therefore ensured by not including $R_{[\mu\nu]}$.}. An important result of this Letter is that there are good reasons to respect projective symmetry and only include the symmetric part of the Ricci because those theories do not exhibit additional dof's, while an explicit breaking of projective symmetry by including the antisymmetric part leads to new pathologies associated to the connection. 

The metric and connection field equations are
\begin{align}
&\frac{\partial F}{\partial g^{\mu\nu}}-\frac{1}{2}F g_{\mu\nu}=T_{\mu\nu},\label{MetricFieldEqs}\\
&\na_\la\lrsq{\sqrt{-q}q^{\mu\nu}}-\del^\mu_\la\na_\rho\lrsq{\sqrt{-q}q^{\rho\nu}}\nonumber\\
&=\sqrt{-q}\lrsq{\cT_{\la\al}^\mu q^{\al\nu}+\cT_{\al\la}^\al q^{\mu\nu}-\del^\mu_\la\cT_{\al\be}^\al q^{\be\nu}},\label{VariationConnection}
\end{align}
where $T_{\mu\nu}=-\frac{2}{\sqrt{-g}}\frac{\delta\cS_{\rm m}}{\delta g_{\mu\nu}}$ is the usual stress-energy tensor, $\cT^\al{}_{\mu\nu}=2\Ga^\alpha{}_{[\mu\nu]}$ is the torsion and we have defined $\sqrt{-q}q^{\mu\nu}\equiv\sqrt{-g}\frac{\partial F}{\partial \cR_{\nu\mu}}$. Since we are assuming that matter fields do not couple to the connection, the corresponding hypermomentum sourcing the connection equation vanishes. Including a non-vanishing hypermomentum will not change our conclusions so we will not consider it here for simplicity \footnote{Including non-minimal couplings in the Ricci-Based theories can be straightforwardly implemented by adding a dependence on $\cR_{\mu\nu}$ in the matter sector. It is easy to see that the only differences will be that Eq. (\ref{def:q1}) will depend on the matter fields and $\cU$ in (\ref{NonSymmetricAction}) will exhibit a more general dependence on the matter fields. A detailed development for the projective invariant case can be found in \cite{Afonso:2017bxr}. }. Although we could work directly with Eqs. (\ref{MetricFieldEqs}) and (\ref{VariationConnection}) in order to understand the number and properties of the dof's, we will do it in a much more transparent manner by going to an Einstein-like frame.

\section{Non-Symmetric Gravity frame}\label{sec:NGTFrame}
 In this section we will show the relation of (\ref{GeneralAction}) with the NGT \cite{Moffat:1978tr}. We start by performing a Legendre transformation of the action (\ref{GeneralAction})  as follows
\beq\label{ModifiedAction}
\cS=\frac12\int \dd^4x \sqrt{-g}\left[F(\Sigma_{\mu\nu})+\frac{\partial F}{\partial \Sigma_{\mu\nu}}\big(\cR_{\mu\nu}-\Sigma_{\mu\nu}\big)\right]
%+\cS_{\rm m}[g,\Psi],
\eeq
where $\Sigma_{\mu\nu}$ is an auxiliary field whose equations of motion $\cR_{\mu\nu}=\Sigma_{\mu\nu}$ can be used to show that \eqref{GeneralAction} and \eqref{ModifiedAction} are on-shell equivalent. We can now perform the field redefinition 
\beq
\sqrt{-q}q^{\mu\nu}=\sqrt{-g}\frac{\partial F}{\partial \Sigma_{\mu\nu}}
\label{def:q1}
\eeq
that gives $\Sigma_{\mu\nu}=\Sigma_{\mu\nu}(\mq,\mg)$, and allows us to express the action as
\beq\label{NonSymmetricActionTwoMetrics}
\cS=\frac12\int \dd^4x\Big[\sqrt{-q}q^{\mu\nu}\cR_{\mu\nu}(\Ga)+\cU(\mq,\mg)\Big]+\cS_{\rm m}[g_{\mu\nu},\Psi],
\eeq
where we have defined the potential
\beq
\cU(\mq,\mg)=\sqrt{-g}\lrsq{F-\frac{\partial F}{\partial\Sigma_{\mu\nu}}\Sigma_{\mu\nu}}_{\Sigma=\Sigma(\mq,\mg)}.
\eeq
We notice now that the metric $g_{\mu\nu}$ enters as an auxiliary field  for minimally coupled matter fields. The field equations $
\frac{\partial \cU}{\partial g^{\mu\nu}}=\sqrt{-g}T_{\mu\nu}$
can then be algebraically solved to obtain $g_{\mu\nu}$ in terms of $q_{\mu\nu}$ and the matter energy-momentum tensor $T_{\mu\nu}$. This solution can be used to integrate $g_{\mu\nu}$ out in \eqref{NonSymmetricActionTwoMetrics} to obtain
\beq\label{NonSymmetricAction}
\cS=\frac12\int \dd^4x \Big[\sqrt{-q}q^{\mu\nu}\cR_{\mu\nu}(\Ga)+\cU(\mq,\mT)\Big]+\cS_{\rm m}[\mg(\mq,\mT),\Psi].
\eeq
It is worth making some comments before proceeding in order to appreciate the crucial differences between the projective and non-projective invariant theories. In theories with projective symmetry, the metric $q^{\mu\nu}$ is symmetric and the $q$ - sector exactly reproduces the first order formulation of GR. Hence, the connection is given by the Levi-Civita connection of $q^{\mu\nu}$, while the matter sector receives new interactions as a consequence of integrating out the space-time metric $g_{\mu\nu}$. The importance of enjoying the projective symmetry lies in that it ensures no new propagating dof's associated to the gravitational sector, and forces the connection to be an auxiliary field that acts as a classical source, generating new matter interactions after being integrated out. We have thus the Einstein frame of these projectively invariant theories (see e.g. \cite{BeltranJimenez:2017doy,Afonso:2017bxr} for a more detailed explanation and also \cite{Delsate:2012ky} where it was already recognised the appearance of new matter interactions within metric-affine gravities). 

The explicit breaking of projective symmetry crucially changes the situation since it translates into the propagation of new dof's that, generally, render the theories unstable. Let us illustrate this by considering vacuum configurations, so the action is given by
\beq
\cS=\frac12\int \dd^4x \Big[\sqrt{-q}\mpl^2q^{\mu\nu}\cR_{\mu\nu}(\Ga)+\cU(\mq)\Big],
\eeq
where we have restored the Planck mass $\mpl$ for convenience. It is then apparent that the vacuum version of these theories reproduces the NGT \cite{Moffat:1978tr} with a potential $\cU$. Former analysis of NGT theories showed that the antisymmetric part of the metric carries a pathological 2-form field that jeopardises their physical viability  \cite{Damour:1991ru,Damour:1992bt}. The instabilities can be seen by considering the antisymmetric sector perturbatively up to quadratic order so that $q_{\mu\nu}=\sq_{\mu\nu}+\frac{\sqrt{2}}{\mpl} (B_{\mu\nu}+\alpha B_{\mu\alpha} B^\alpha{}_\nu+\beta B^2\sq_{\mu\nu})$, with $\sq_{\mu\nu}$ an arbitrary symmetric metric, $B_{\mu\nu}$ a 2-form field that encodes $q_{[\mu\nu]}$, and where the parameters $\alpha$ and $\beta$ reflect the possibility of field redefinitions at quadratic order. When expanding around such a background at second order in $B_{\mu\nu}$ we have \cite{Damour:1992bt}:
\begin{align}
\cS^{(2)}=&\int\dd^4x\sqrt{-\sq}\left[\frac12\mpl^2R(\sq)-\frac{1}{12} H^2-\frac14 m^2 B^2\right.\nonumber\\
&-\frac{\sqrt{2}\mpl}{3}B^{\mu\nu}\partial_{[\mu}\Gamma_{\nu]}+\frac14\big(1-2\alpha+4\beta\big) R(\sq) B^2\nonumber\\
&+\alpha R_{\mu\nu}(\sq)B^{\mu\alpha}B^\nu{}_\alpha-R_{\mu\nu\alpha\beta}(\sq)B^{\mu\alpha}B^{\nu\beta}\Big]
\label{eq:Actquad}
\end{align}
where $H^2$ is the usual 2-form field kinetic term, $m^2$ the mass generated by $\cU$, and $\Gamma_\mu$ is the projective mode of the connection. It has been argued that the mass can cure some pathologies associated to the curvature couplings in NGT \cite{Damour:1992bt}, but some instabilities persist \cite{Clayton:1996dz}. 

In order to show the ghostly nature of the projective mode, we will first consider a maximally symmetric background with $R_{\mu\nu\alpha\beta}(\sq)=\Lambda(\sq_{\mu\alpha}\sq_{\nu\beta}-\sq_{\mu\beta}\sq_{\nu\alpha})$ and a frozen $\sq_{\mu\nu}$ so the non-minimal couplings simply amount to a change in the mass $m^2\rightarrow\tilde{m}^2(\Lambda)$. Around a flat background, $\Lambda=0$, the mass remains $m^2$. Then, we can diagonalise the action with the field redefinition $B_{\mu\nu}=\tilde{B}_{\mu\nu}-\frac{2\sqrt{2}\mpl}{3\tilde{m}^2}\partial_{[\mu}\Gamma_{\nu]}$. Since this redefinition has the form of a gauge transformation, the kinetic term of the 2-form remains unaffected and the action reads
\begin{align}\label{ActionDec}
\cS^{(2)}_{\rm flat}=\int\dd^4x\sqrt{-\sq}\left(-\frac{1}{12} H^2-\frac14 \tilde{m}^2 \tilde{B}^2+\frac14 \partial_{[\mu}\tilde{\Gamma}_{\nu]}\partial^{[\mu}\tilde{\Gamma}^{\nu]} \right)
\end{align}
where we have introduced the canonically normalised field $\tilde{\Gamma}_\mu=\frac{2\sqrt{2}\mpl}{3\tilde{m}}\Gamma_\mu$. The wrong sign for the kinetic term of the projective mode clearly shows the presence of a massless spin-1 ghost in the projective sector that signals the presence of a fatal instability.

After showing how the projective mode propagates a ghost around maximally symmetric and fixed backgrounds, let us show how the couplings to the curvature when $\sq_{\mu\nu}$ is unleashed in (\ref{eq:Actquad}) present additional pathologies, which have also been discussed for NGT in \cite{Damour:1992bt}. The nature of these pathologies can be interpreted as Ostrogradski instabilities \cite{Woodard:2015zca} associated to having higher order equations of motion\footnote{The Ostrogradski instabilities have not been properly identified within NGT and represent yet another problem for NGT besides the pathological asymptotic behaviour diagnosed in \cite{Damour:1992bt}.}. One way of understanding the instabilities is by noticing that, after diagonalising, the field strength of the projective mode will couple to the curvature. An alternative procedure is recalling that a massive 2-form field can be dualised to  a massive vector field in 4 dimensions. The dualisation is such that the field strength of the 2-form is dualised to a vector field as $H^{\alpha\beta\gamma}=\frac16 \epsilon^{\alpha\beta\gamma\mu} A_\mu$, while the 2-form field becomes the dual of the field strength $F_{\alpha\beta}=2\partial_{[\alpha} A_{\beta]}$, i.e., $B^{\mu\nu}=\frac12\epsilon^{\mu\nu\alpha\beta}F_{\alpha\beta}$.  Thus, the couplings to the curvature of the 2-form will give rise to non-minimal interactions for the vector field with the schematic form $\sim RFF$ in the dual representation\footnote{The same conclusion is reached by introducing Stueckelberg fields $b_{\mu}$ and taking the appropriate decoupling limit. Then, the Stueckelberg fields feature analogous non-minimal couplings for their field strength \cite{affineghosts}, giving a third view on the problem.}. Either way, we see the appearance of couplings between curvature and field strengths. Unless these couplings precisely correspond to the Horndeski vector-tensor interactions \cite{Horndeski:1976gi}, it is well-known that they lead to higher order field equations and, consequently, prone to Ostrogradski instabilities. It is interesting to note that one can in fact reach the healthy Horndeski interaction by means of a field redefinition at quadratic order (with an appropriate choice of $\alpha$ and $\beta$). However, even in this case, pathologies around important cosmological and astrophysical backgrounds arise \cite{Jimenez:2013qsa}. Furthermore, the non-existence of healthy higher dimension operators involving curvatures and $F_{\mu\nu}$ in 4 space-time dimensions \cite{Horndeski:1976gi,Jimenez:2016isa} signals that higher order terms in $B_{\mu\nu}$ will reintroduce the Ostrogradski instabilities. Let us note that, while the ghost around maximally symmetric backgrounds can be easily cured by adding  $\Gamma_{\mu\nu} \Gamma^{\mu\nu}$ (permitted if we allow for a more general framework beyond RBG), the ghosts associated to the non-minimal couplings are more difficult to evade, if possible at all.\\

Our discussion clearly shows the presence of five propagating fields contained in the connection (the three dof's of the massive 2-form plus the two polarisations of the projective mode), in sharp contrast to the projectively invariant case where there are no new propagating modes. It is precisely these fields, arising from the explicit breaking of projective symmetry, that root the pathologies present in general RBG theories. Notice that the higher order nature of the equations makes the 2-form field propagate more than the expected three modes and these additional modes are in turn carry the Ostrogradski instability.

Finally, let us emphasise that these instabilities arise already in the gravitational sector without including matter. However, as explained above, new interactions in the matter sector will be generated after integrating out $g_{\mu\nu}$, and in particular, matter couplings to $B_{\mu\nu}$ that could introduce yet additional pathologies. Similarly, had we considered direct couplings of the connection to matter, the same conclusion would be reached for vacuum configurations \cite{affineghosts}.

%Now that we have shown the presence of ghosts in the general RBG theories we can ask if we can render the theories stable with some mechanism. We have already shown on possible solution that consists in imposing a projective symmetry. In the following, we will show an alternative solution based on the introduction of appropriate constraints.

\section{Exorcising the ghosts: Torsion-free theories}

So far we have seen how the projective symmetry is of paramount importance to avoid ghost-like instabilities in RBG. We will show now how to avoid such instabilities without imposing projective symmetry, but rather constraining the theory to be torsion-free. This can be easily implemented by adding suitable Lagrange multiplier fields enforcing $\cT^\alpha{}_{\mu\nu}=0$ so the connection field equations are
\beq\label{ConnectionFieldEqsTorsionFree}
\na_\la\lrsq{\sqrt{-g}f^{(\mu\nu)}}-\na_\rho\lrsq{\sqrt{-g}f^{\rho(\mu}}\del^{\nu)}_\la=0,
\eeq
where we have introduced $f^{\mu\nu}\equiv\partial f/\partial \cR_{\mu\nu}$. Let us decompose it as $
\sqrt{-g}f^{\mu\nu}=\sqrt{-h}h^{\mu\nu}+\sqrt{-h}B^{\mu\nu}$
with $h^{\mu\nu}\equiv f^{(\mu\nu)}$ and $B^{\mu\nu}\equiv f^{[\mu\nu]}$ the symmetric and antisymmetric parts respectively. Since the torsion is constrained to vanish, we can conveniently decompose the connection in terms of the Levi-Civita of $h^{\mu\nu}$ and a disformation part $L^\alpha{}_{\mu\nu}$ as
\beq\label{ConnectionDeccomposition}
\Ga_{\mu\nu}^\al=\bar{\Ga}_{\mu\nu}^\al(h)+L^\al{}_{\mu\nu}.
\eeq
%The above splitting allows to obtain the following relations that we will use below
%\begin{align}
%\nabla_\lambda\big(\sqrt{-h}h^{\lambda\nu}\big)&=\sqrt{-h}\bL^\nu,\label{L1}\\
%\nabla_\lambda\big(\sqrt{-h}B^{\lambda\nu}\big)&=\sqrt{-h}\bar{\nabla}_\lambda B^{\lambda\nu},\label{L2}
%\end{align}
%where $\bL^\nu\equiv L^\nu{}_{\alpha\beta} h^{\alpha\beta}$ is one of the two independent traces of the disformation tensor. 
Using this decomposition in the connection equation (\ref{ConnectionFieldEqsTorsionFree}), its trace and its contraction with $h_{\mu\nu}$, defined as the inverse of $h^{\mu\nu}$, lead to  
\begin{eqnarray}
\bar{\nabla}_\lambda B^{\lambda \nu}&=&\frac{1-D}{1+D}\bL^\nu, \label{eq:constraint1}\\
L_\mu&=&\frac{2}{(2-D)(1+D)}\bL^\alpha h_{\alpha \mu}\label{eq;Tracesconst},
\end{eqnarray}
where $L_{\mu}\equiv L^\alpha{}_{\mu\alpha}$ and $\bL^\nu\equiv L^\nu{}_{\alpha\beta} h^{\alpha\beta}$ are the two traces of the disformation tensor, which are in turn dynamically related by (\ref{eq;Tracesconst}). Eq. ({\ref{eq:constraint1}) however implies the transversality constraint
\beq
\bar{\nabla}_\nu \bL^\nu=0.
\label{eq:constraint2}
\eeq
When inserting the above relations into the connection equation (\ref{ConnectionFieldEqsTorsionFree}), we arrive at
\beq
2h^{\alpha(\mu} L^{\nu)}{}_{\lambda\alpha}=L_\lambda h^{\mu\nu}+(2-D) L_\alpha h^{\alpha(\mu}\delta^{\nu)}{}_\lambda.
\label{eq:connectionTfree2}
\eeq
We need to recall now the definition of the non-metricity tensor 
$Q_\lambda{}^{\mu\nu}\equiv -\nabla_\lambda h^{\mu\nu}=-2h^{\alpha(\mu}L^{\nu)}{}_{\lambda\alpha}$
which also gives the relation $L_\mu=-\frac12h_{\alpha\beta}Q_\mu{}^{\alpha\beta}\equiv-\frac12 \tilde{Q}_\mu$. These relations allow us to express (\ref{eq:connectionTfree2}) as
\beq\label{relationtracesL}
Q_\lambda{}^{\mu\nu}=\frac12\Big[ \tilde{Q}_\lambda h^{\mu\nu}+(2-D)\tilde{Q}_\alpha h^{\alpha(\mu}\delta^{\nu)}_\lambda\Big].
\eeq
We have then solved for the full connection as the Levi-Civita of $h^{\mu\nu}$ plus a disformation part determined by the above non-metricity tensor. We see that the non-metricity is fully determined by its trace so that there is only one additional vector field associated to the connection. Furthermore, from  the constraint (\ref{eq:constraint2}) we conclude that this vector field propagates 3 degrees of freedom, corresponding to a Proca field. The resolution of the problem will then be completed by considering the Einstein equations, which allow to algebraically solve for $h^{\mu\nu}$ in terms of the matter fields (possibly including the vector $\tilde{Q}_\mu$). A particular case was considered in \cite{Buchdahl:1979ut,Vitagliano:2010pq} for $f\propto\cR+c_1 \cR_{[\mu\nu]}\cR^{[\mu\nu]}$, where it was shown that this action exactly reproduces the Proca Lagrangian for the connection sector. In the more general case under consideration here, there will be more involved interactions for the Proca field, as it was also found in \cite{Olmo:2013lta}.

We can gain some clearer intuition by reformulating these theories in the Einstein frame. For that, let us rewrite our original action as follows:
\begin{align}\label{NoTorsionEinstein}
\cS=&\frac12\int\dd^Dx\sqrt{-g}\Big[f(\Sigma,A)+\frac{\partial f}{\partial \Sigma_{\mu\nu}}\big(\cR_{(\mu\nu)}-\Sigma_{\mu\nu}\big)+
\nonumber\\
&+\frac{\partial f}{\partial A_{\mu\nu}}\big(\cR_{[\mu\nu]}-A_{\mu\nu}\big)+\frac{1}{\sqrt{-g}}\lambda_\alpha{}^{\mu\nu}\cT^\alpha{}_{\mu\nu}\Big]
\end{align}
where $\Sigma_{\mu\nu}$ and $A_{\mu\nu}$ are two symmetric and antisymmetric auxiliary fields respectively and $\lambda_\alpha{}^{\mu\nu}$ is a Lagrange multiplier field enforcing $\cT^\alpha{}_{\mu\nu}=0$. We can again perform field redefinitions analogous to those in Sec. \ref{sec:NGTFrame}, and integrate out the space-time metric $g_{\mu\nu}$. After doing that, \eqref{NoTorsionEinstein} becomes
\begin{align}
\cS=&\frac12\int\dd^Dx\Big[\sqrt{-h}h^{\mu\nu}\cR_{(\mu\nu)}+\sqrt{-h}B^{\mu\nu}\cR_{[\mu\nu]}\nonumber\\
&+\sqrt{-h}\cU(h,B,T)+\lambda_\alpha{}^{\mu\nu}\cT^\alpha{}_{\mu\nu}\Big].
\end{align}
The connection equations for this action are formally the same as \eqref{ConnectionFieldEqsTorsionFree}, so we can simply take the solution for the connection, essentially the splitting (\ref{ConnectionDeccomposition}) with the solution (\ref{relationtracesL}),  and insert it into the action. Since the solution for the connection satisfies
\begin{align}
\cR_{(\mu\nu)}=&R_{\mu\nu}(h)+\frac{(D-2)(D-1)}{16}\tilde{Q}_\mu \tilde{Q}_\nu\\&-\frac{(D-1)}{4}h_{\mu\nu}\bar{\nabla}_\alpha \tilde{Q}^\alpha\nonumber \\
\cR_{[\mu\nu]}=&-\frac{1}{2}\partial_{[\mu}\tilde{Q}_{\nu]},
\end{align}
our final action can be expressed as
\begin{align}
\begin{split}
\cS=\frac12\int\dd^Dx\sqrt{-h}\Big[&R(h)+\frac{(D-2)(D-1)}{16}\tilde{Q}^2\\
&-\frac{1}{2}B^{\mu\nu}\partial_{[\mu}\tilde{Q}_{\nu]}+\cU(h,B,T)\Big],
\end{split}
\label{eq:finalaction}
\end{align}
where we have dropped the boundary term $\bar{\nabla}_\mu \tilde{Q}^\mu$. Notice that this form of the action reproduces (\ref{eq:constraint1}) as\beq
\bar{\nabla}_\mu B^{\mu\nu}=-\frac{(D-2)(D-1)}{4}\tilde{Q}^\nu,
\eeq
which recuperates the constraint $\bar{\nabla}_\alpha \tilde{Q}^\alpha=0$. On the other hand, the equation for $B^{\mu\nu}$ yields
\beq
\partial_{[\mu}\tilde{Q}_{\nu]}=2\frac{\partial{\cU}}{\partial B^{\mu\nu}}
\eeq
which gives the (non-linear) relation between the field strength of $\tilde{Q}_\mu$ and the 2-form $B^{\mu\nu}$, also involving the matter fields. This is a reflection of the fact that our final action (\ref{eq:finalaction}) is the first order form of a massive vector field with self-interactions and couplings to the matter fields. We can easily reproduce the result in \cite{Buchdahl:1979ut,Vitagliano:2010pq} for $f\propto\cR+c_1 \cR_{[\mu\nu]}\cR^{[\mu\nu]}$. In that case, the metric $h^{\mu\nu}$ is directly $g^{\mu\nu}$, while the effective potential reduces to $\cU\propto B^2$ so that (\ref{eq:finalaction}) exactly reproduces the first order form of a free Proca field $\tilde{Q}_\mu$. The same result was found in \cite{Olmo:2013lta} for theories built with the Ricci-squared scalar, and we have reached here the same conclusion for a general RBG with vanishing torsion in a more explicit form.

\section{Hybrid theories}
In order to give a more complete discussion of RBG, we will finally consider them within the hybrid framework \cite{Harko:2011nh, Capozziello:2015lza}, described by the action
\beq
\cS_{\rm hybrid}=\int\dd^Dx\sqrt{-g}f(\cR_{\mu\nu},R_{\mu\nu})
\eeq
where $\cR_{\mu\nu}=\cR_{\mu\nu}(\Gamma)$ and $R_{\mu\nu}=R_{\mu\nu}(g)$ are the Ricci tensors of the affine connection and the Levi-Civita connection of $g_{\mu\nu}$ respectively. The general pathologies exhibited by these theories can be straightforwardly identified by going to their bimetric formulation (see \cite{Koivisto:2013kwa} for a discussion on pathologies of hybrid theories). Having a non-linear dependence with $R_{\mu\nu}(g)$ already introduces ghosts, so we will restrict the metric sector to the Einstein-Hilbert term and will focus on actions of the form
\beq
\cS_{\rm hybrid}=\int\dd^Dx\sqrt{-g}\left[\frac12 R(g)+f(\cR_{\mu\nu})\right]
\eeq
that will suffice to illustrate the problems with these theories. We can then follow the same procedure as above for the affine sector by writing the hybrid action in the bimetric form
\beq
\cS_{\rm hybrid}=\int\dd^Dx\left[\frac{\sqrt{-g}}{2} R(g)+\frac{\sqrt{-q}}{2}q^{\mu\nu}\cR_{\mu\nu}(\Gamma)+\cU(q,g)\right],
\label{eq:hybridbimetric}
\eeq
which resembles (\ref{NonSymmetricActionTwoMetrics}), but it presents some crucial differences that make it even more pathological. If we take the decoupling limit of the $g$-sector (technically by sending the corresponding Planck mass to infinity), we would still have the NGT sector with the same problems. However, the hybrid theories are generally pathological even if the projective symmetry is imposed on the affine sector so that $q_{\mu\nu}$ is a symmetric metric. In that case, the action (\ref{eq:hybridbimetric}) describes a bimetric theory with an interaction potential given by $\cU(g,q)$ (see also \cite{BeltranJimenez:2017doy}). As it is well-known, only a very specific tuning of the potential allows to remove the Boulware-Deser ghost \cite{Boulware:1973my} of these theories \cite{bimetric} and, consequently, Ricci-based hybrid theories are even more prone to instabilities than their metric-affine formulation. The bi-metric construction fails for theories of the type $f(R,\cR)$ so our conclusion does not apply to them (see however \cite{Koivisto:2013kwa} for pathologies of those theories as well).

\section{Discussion}

We have shown that general RBG theories suffer from ghost-like instabilities in the additional dof's associated to the connection and  which arise from the explicit breaking of projective symmetry. Having the projective symmetry then proves to be crucial for the viability of RBG, in which case the theories reduce to GR with some new matter interactions. Additionally, we have shown that the projective symmetry is not required if the connection is constrained to be torsion-free and the theory then contains one additional massive vector field. We have extended our discussion to the hybrid framework where, even with a projective symmetry, the theories typically propagate a Boulware-Deser ghost.\\

It is worth emphasising that, although we have only considered RBGs, our results extend to general metric-affine theories, since including more geometrical objects in the action will typically introduce even more potentially unstable propagating modes.  Let us stress however that there will be non-pathological higher order curvature theories, like e.g. theories for which the metric and metric-affine formalisms are equivalent \cite{Exirifard:2007da,Borunda:2008kf}, but the results presented in this Letter clarify that resorting to the metric-affine formalism for higher order curvature theories does not, in general, guarantee the absence of ghosts, thus sharing analogous pathologies with the metric approach. In this respect, one needs to be cautious when considering higher order curvature theories in the metric-affine formalism (by imposing symmetries and/or constraints), similarly to the metric framework where only judicious combinations of curvatures like the Lovelock terms lead to physically sensible theories.

%%%%%%%%%%%%%%%%%%%%%%%%%%%
\acknowledgments 
We would like to thank Dario Bettoni, Tomi S. Koivisto, Diego Rubiera-Garc\'ia and Gonzalo J. Olmo for useful discussions and comments. JBJ acknowledges support from the  {\textit{ Atracci\'on del Talento Cient\'ifico en Salamanca}} programme and the MINECO's projects FIS2014-52837-P and FIS2016-78859-P (AEI/FEDER). AD is supported by a PhD contract of the program FPU 2015 (Spanish Ministry of Economy and Competitiveness) with references FPU15/05406. This work is supported by the Spanish project FIS2014-57387-C3-1-P (MINECO/FEDER, EU), the project H2020-MSCA-RISE-2017 Grant FunFiCO-777740, the project SEJI/2017/042 (Generalitat Valenciana), the Consolider Program CPANPHY-1205388, and the Severo Ochoa grant SEV-2014-0398 (Spain). This article is based upon work from COST Action CA15117, supported by COST (European Cooperation in Science and Technology). AD also wants to thank hospitality to the \textit{Departamento de F\'isica Te\'orica de la Universidad de Salamanca}.
%%%%%%%%%%%%%%%%%%%%%%%%%%%

%\bibliography{Bibliography}
%\bibliographystyle{unsrt}

\end{document}